# Caracterización del paisaje y análisis del límite oeste del Parque Nacional Carara, Costa Rica: implicaciones para la conservación de la biodiversidad

Sergio Gabriel Quesada-Acuña, Valerie Valdelomar, Cristina Arrieta-Ch., Gustavo Ruiz Morales, Miguel Matarrita-Herrera, José Fabio Araya G. & Iván Sandoval Hernández

Escuela de Ciencias Biológicas, Universidad Nacional, Apdo. 86-3000, Heredia, Costa Rica. sgbiotropic@gmail.com

**ABSTRACT.** It is unknown if the current size and shape of protected areas (PA) in Costa Rica favors retention of viable populations of wildlife. The western border of Carara National Park (CNP) and its surroundings were characterized at a landscape level in order to consider their implications for biodiversity conservation. 54 400 hectares of Costa Rican Central Pacific tree cover were analyzed (1997, 2000, 2005) to calculate the size of the PAs, mean shape index, mean patch fractal dimension and edge density. Aside from that, we performed eight habitat evaluations on the western border of the CNP ("costanera-sur" highway) to assess tree composition, regeneration, litter, horizontal obstruction, and canopy cover. It was determined that the PAs were between 36 and 5 242 hectares long. We observed that the tree cover increased from 21 231,8 hectares in 1997 to 29 006,9 hectares in 2000, and decreased to 26 933,4 hectares in 2005. We found out that most of the tree cover (2005) belongs to CNP and ZP Cerros de Turrubares, but both of them present high values of edge density and mean shape index, suggesting that they are susceptible to alteration and fragmentation. The four evaluated points have a similar successional stage. In order to maintain the potential of CNP as an area for biodiversity conservation, we recommend reducing the maximum speed limit in the region adjacent to the park. It is also important to establish frequent monitoring of the tree cover and promote reforestation programs to create corridors that stop the area's fragmentation and facilitate biodiversity conservation.

**RESUMEN.** Se desconoce si el tamaño y forma actual de las áreas protegidas (AP) en Costa Rica favorece la permanencia de poblaciones viables de vida silvestre. Se caracterizó a nivel de paisaje el límite oeste del PN Carara y alrededores, para considerar sus implicaciones en la conservación de la biodiversidad. Se analizó la cobertura forestal (1997, 2000, 2005) de 54 400 ha en el Pacífico Central de Costa Rica; para calcular tamaño de las AP, índice de complejidad de forma, índice de dimensión fractal y densidad de bordes. Además se hicieron ocho evaluaciones de hábitat en el límite oeste del PNC (carretera costanera-sur) para evaluar composición arbórea, regeneración, hojarasca, obstrucción horizontal y cobertura del dosel. Se determinó que las AP miden entre 36 y 5 242 ha. Se observó que la cobertura forestal de 1997 (21 231,8 ha), aumentó para el año 2000 (29 006,9 ha) y disminuyó para el año 2005 (26 933,4 ha). En la cobertura forestal (2005) se comprobó que la mayor parte de los terrenos protegidos pertenecen al PNC y la ZP cerros de Turrubares, pero ambas presentan valores altos de densidad de bordes y complejidad de formas, lo cual sugiere que son susceptibles a alteración y fragmentación. Los cuatro puntos evaluados presentan un estado sucesional similar. Para mantener el potencial del PNC como área de conservación de la biodiversidad, es recomendable reducir el límite de velocidad máxima en la región que colinda con el PNC. También es importante establecer un monitoreo frecuente de la cobertura boscosa y fomentar programas de reforestación para crear corredores biológicos que detengan la fragmentación de la zona y faciliten la conservación de la biodiversidad.

La reducción y fragmentación del hábitat es considerada una de las principales causas de extinción de la biodiversidad mundial, ya que la presencia y establecimiento de muchas especies de plantas y animales, depende del tamaño y forma de los fragmentos de bosque (Harris 1984, Beier 1993, Simonetti & Mella 1997, Rau & Gantz 2001, Pincheira *et al.* 2009, Gurrutxaga-San Vicente & Lozano-Valencia 2010).

Entre 1970 y 1980, la deforestación en Costa Rica redujo entre un 66% y 80% la cobertura boscosa, principalmente en el norte del país y en las tierras bajas de ambas vertientes (Harrison 1991, Carrillo & Vaughan 1994, Bonilla-Carrión & Rosero-Bixby 2004), lo cual provocó la fragmentación del paisaje y el aislamiento de las Áreas Protegidas (AP), con la consecuente pérdida de hábitats naturales y reducción de la biodiversidad (Troche & Guarachi 2001, Martínez-Salinas 2008). Este escenario se mantiene en la actualidad, pues los pequeños fragmentos protegidos por lo general se encuentran aislados, rodeados por un paisaje de matrices antrópicas continuas, generando AP semejantes a islas de diferentes formas y tamaños (Preston 1960, Diamond & May 1976, Harris 1984, Rau & Gantz 2001).

Desde 1970 se han establecido en Costa Rica, diversas AP, como estrategia para mitigar los efectos de la fragmentación, las cuales actualmente se encuentran amenazadas por procesos de aislamiento y reducción del hábitat (Harrison 1991, Schelhas & Pfeffer 2009). Se sabe que a mayor tamaño de un parche, es posible mantener una mayor heterogeneidad espacio-temporal que favorezca la biodiversidad (Pickett & Thompson 1978), pero se desconoce si el estado actual de las AP en cuanto a tamaño y forma favorece la biodiversidad y la permanencia de poblaciones viables de vida silvestre.

El Parque Nacional Carara (PNC), al igual que el resto de AP del país, se ha visto amenazado por el cambio de uso del suelo en sus alrededores. Creado en 1978 como Reserva Biológica, fue nombrado en su categoría actual en 1998, como un intento de proteger el último remanente de bosque transicional seco-húmedo del país (Laurencio & Malone 2009); característica que lo convierte en una "isla biológica" y un refugio obligatorio para la fauna desplazada por el intenso uso agropecuario, el cambio de uso del suelo y la fragmentación del hábitat que enfrenta la zona (Ávila 2002, Sandoval-Hernández 2003).

Este trabajo se realizó con el objetivo de caracterizar el paisaje alrededor del PNC y principalmente su límite oeste (carretera costanera-sur, ruta 34), analizando las implicaciones que presenta este sitio para la conservación de la biodiversidad.

## MATERIAL Y MÉTODOS

**Sitio de estudio**: El Parque Nacional Carara (9°45'22" N, 84°36'27" W) es un área protegida de 5 242 ha, ubicada en Puntarenas, Pacífico Central de Costa Rica, donde su límite norte es el río Tárcoles y su límite oeste es la carretera costanera-sur (ruta 34). El 95% de la cobertura del PNC corresponde a bosques primarios, secundarios y de galería, pero también es posible encontrar pantanos y bosques inundables (Vargas 1992). La elevación en el PNC varía entre 30-636 m, la precipitación alcanza 2000-3000 mm anuales según la época del año, y la temperatura anual es de 27,8 °C (Vargas 1992).

Se analizó a nivel de paisaje, el tipo de cobertura de 54 400 hectáreas (ha) en el Área de Conservación Pacífico Central (ACOPAC); tomando como núcleo el PNC y abarcando las AP más cercanas (Refugio de Vida Silvestre Fernando Castro, RVS Finca Hacienda La Avellana, RVS Surtubal, RVS Cacyra y Zona Protectora Cerros de Turrubares) (Fig. 1). Para el análisis se utilizó el programa ArcView GIS 3.3, la extensión Patch Analyst (Elkie *et al*. 1999) y las bases de datos de cobertura y áreas protegidas (AP) de los años 1997, 2000 y 2005, pertenecientes al Atlas Digital de Costa Rica del Instituto Tecnológico de Costa Rica (Ortiz & Soto 2008). Para cada AP se calculó la cobertura forestal (año 2005) y los índices de forma de los fragmentos, incluyendo: Densidad de bordes (DB): Medida en metros por hectárea (m/ha) de la cantidad relativa de borde en el área. Índice complejidad de formas (CF): donde CF=1 si los parches son circulares o cuadrados. Índice de dimensión fractal (DF): donde valores cercanos a 1 indican formas euclidianas y valores cercanos a 2 indican formas fractales. Posteriormente se analizó la cobertura forestal total del área de estudio con las mismas variables para cada año.

Además se hicieron ocho evaluaciones de hábitat en cuatro sitios del límite oeste del PNC (carretera costanera-sur), de los cuales tres se ubicaron dentro del PNC (laguna meándrica, centro de visitantes y quebrada bonita) y uno

**Cuadro 1.** Tamaño e índices de forma de las distintas áreas protegidas dentro del área estudiada en el ACOPAC, Costa Rica.

| Categoría | Nombre | Extensión (ha) | DB | CF | DF |
|---|---|---|---|---|---|
| PN | Carara | 5 242,24 | 4,88 | 1,93 | 1,22 |
| ZP | Cerros de Turrubares | 2 867,85 | 3,77 | 2,02 | 1,23 |
| RVS | Finca Hacienda La Avellana | 508,32 | 1,18 | 1,50 | 1,22 |
| RVS | Fernando Castro Cervantes | 191,43 | 3,90 | 2,42 | 1,27 |
| RVS | Surtubal | 136,19 | 0,56 | 1,38 | 1,22 |
| RVS | Cacyra | 36,68 | 0,26 | 1,23 | 1,23 |

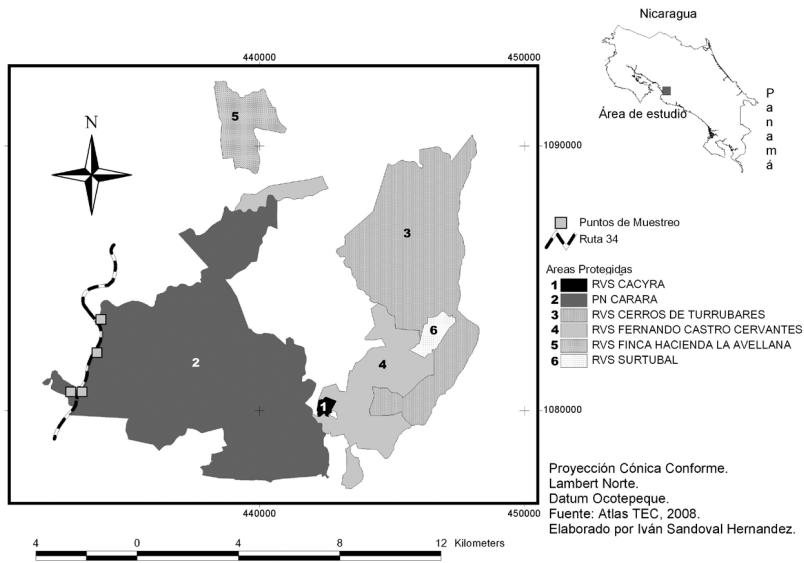

**Figura 1.** Sitio de estudio en ACOPAC, Costa Rica. Se observa el Parque Nacional Carara y alrededores, con los cuatro puntos de muestreo a lo largo de la carretera costanera-sur, ruta 34.

fuera del PNC (parche boscoso privado), en el único parche aledaño de bosque remanente (Fig. 1). En cada sitio se establecieron dos parcelas de 50 x 50 m, subdivididas en cuatro cuadrantes de 25 x 25 m, entre los cuales se midió: Composición arbórea: se anotó la altura, especie y diámetro a la altura del pecho (DAP) de todos los árboles con DAP mayor o igual a 10 cm. Regeneración: se contaron todos los árboles y arbustos con DAP menor a 10 cm y entre 1-3 m de altura. Hojarasca: Se utilizó una regla de metal y se hicieron cinco mediciones de profundidad de hojarasca para obtener un promedio. Obstrucción horizontal: Se hicieron tres estimaciones utilizando una manta

**Cuadro 2.** Cambio en la cobertura forestal e índices de forma en años anteriores, dentro del área estudiada en el ACOPAC, Costa Rica.

| Año | Cobertura forestal (ha) | DB | CF | DF |
|---|---|---|---|---|
| 1997 | 21 231,86 | 20,82 | 13,07 | 1,40 |
| 2000 | 29 006,99 | 13,71 | 14,32 | 1,40 |
| 2005 | 26 933,49 | 18,68 | 13,46 | 1,40 |

**Cuadro 3.** Promedios de hojarasca, obstrucción horizontal, coberturas de dosel y árboles de regeneración, para cuatro puntos de muestreo en el límite oeste del PNC, ACOPAC, Costa Rica.

| Punto | Profundidad de Hojarasca (mm) | Obstrucción Horizontal (%) | Cobertura del dosel (%) | Regeneración (individuos) |
|---|---|---|---|---|
| Lag. meándrica | 41,10 | 46,67 | 97,27 | 87 |
| C. de visitantes | 67,44 | 73,30 | 97,87 | 482 |
| Quebrada bonita | 35,50 | 60,83 | 97,45 | 219 |
| Parche boscoso | 79,25 | 55,50 | 92,49 | 272 |

**Cuadro 4.** Número total de árboles por categoría de tamaño de DAP y por sitio de muestreo, en el límite oeste del PNC (carretera costanera-sur, ruta 34), ACOPAC, Costa Rica.

| Categoría DAP | Lag. meándrica | C. de visitantes | Quebrada Bonita | Parche boscoso |
|---|---|---|---|---|
| 11-20 cm | 14 | 13 | 17 | 6 |
| 21-30 cm | 8 | 7 | 6 | 3 |
| 31-40 cm | 4 | 2 | 4 | 2 |
| Más de 40 cm | 11 | 7 | 6 | 10 |
| Totales | 37 | 29 | 33 | 21 |

blanca de 50x80 cm ubicada a 10 m de un mismo observador. Cobertura del dosel: Utilizando un densitómetro esférico (Modelo A) se estimó la cobertura 20 veces por parcela para obtener un promedio. Las distintas variables estudiadas fueron analizadas con las pruebas Kruskall-Wallis, chi cuadrado simple y rangos múltiples, utilizando el programa Statgraphics Centurion XVI (StatPoint Technologies, Inc).

## RESULTADOS

Las AP estudiadas según la cobertura 2005, tienen tamaños variables entre 36 ha (RVS Cacyra) y 5 242 ha (PNC). Los mayores valores para densidad de bordes (DB) e índice de complejidad de formas (CF) se encuentran en el PNC, RVS Fernando Castro y ZP Cerros de Turrubares, mientras que la dimensión fractal (DF) es similar en todas las AP (Cuadro 1).

La cobertura forestal de la zona para el año 1997 (21 231,8 ha), aumentó de manera considerable para el año 2000 (29 006,9 ha) y disminuyó al 2005 (26 933,4 ha). Los valores para densidad de borde (DB) disminuyen conforme aumenta la cobertura forestal; mientras que el índice de complejidad de formas (CF) presentó valores altos (13-14) y la dimensión fractal (DF) valores relativamente bajos (1,4) en los tres años evaluados (Cuadro 2).

Las evaluaciones del hábitat en el límite oeste del PNC, mostraron que la cantidad de hojarasca es diferente entre los sitios (KW= 18,777; P= 0,0003), siendo mayor en el parche boscoso frente al PNC (7,93 cm) y menor en quebrada Bonita (3,55 cm) (Cuadro 3).

Los valores de obstrucción horizontal no fueron diferentes entre los sitios (KW= 4,709; P= 0,194); pero si se observó diferencia significativa entre las coberturas de dosel de los sitios (KW= 71,181; P= 0,0), donde el parche boscoso frente al PNC obtuvo el menor valor (92,49%). También se observó diferencia entre la cantidad de árboles jóvenes (regeneración) para cada sitio, donde el centro de visitantes y el parche boscoso obtuvieron la mayor cantidad (Cuadro 3).

En cuanto a composición arbórea, la laguna meándrica presentó la mayor cantidad de árboles (37), seguida por quebrada bonita (33), el centro de visitantes (29) y el parche boscoso frente al PNC (21) (Cuadro 4). Las especies de árboles más frecuentes en los distintos puntos fueron *Anacardium excelsum, Bravaisia integerrima, Ceiba pentandra, Clarisia biflora, Crateva tapia, Ficus maxima* y *Quararibea asterolepis*.

## DISCUSIÓN

La reducción y aislamiento de las áreas protegidas es una de las principales amenazas para la biodiversidad, pues se ha demostrado que generan fragmentación de ecosistemas, pérdida de hábitats, incremento del efecto de borde, reducción de las poblaciones silvestres, homogenización de la composición de especies, alteración de la dinámica y microclimas del bosque, entre otros (Bennett 1990, De lima & Gascon 1999, Schelhas & Pfeffer 2009, Gurrutxaga-San Vicente & Lozano-Valencia 2010). Este tipo de procesos generados por la intervención humana, provocan que las AP tiendan a ser pequeñas e irregulares, y en ellas la biodiversidad comienza a descender drásticamente, como lo explica la teoría del aislamiento geográfico aplicada a paisajes fragmentados (Diamond & May 1976; Harris 1984, Caughley & Sinclair 2002).

En este trabajo, al analizar la cobertura forestal de 54 400 ha del ACOPAC (2005), se comprobó que la mayor parte de los terrenos protegidos pertenecen al PNC y la ZP cerros de Turrubares (Cuadro 1), lo cual es importante porque permite que los procesos naturales se mantengan en el tiempo, conservando mayor cantidad de especies y poblaciones más estables (Pickett & Thompson 1978, Bennett 1990). Sin embargo, también se encontró que los mayores valores para densidad de bordes (DB) corresponden a éstos mismos sitios (Cuadro 1), lo cual indica que son áreas expuestas a efectos de borde y susceptibles a la alteración, la fragmentación y el cambio de las condiciones medioambientales (Bennett 1990, Guariguata & Kattán 2002).

Del mismo modo, el índice de complejidad de formas (CF) resultó mayor precisamente en las AP de mayor tamaño (Cuadro 1), indicando que éstas poseen formas irregulares, complejas y alargadas, que por su perímetro dentado son propensas a fragmentación por el llamado "efecto península", el cual explica que las áreas alargadas suelen ser alteradas en su anchura menor hasta que son divididas en dos partes más pequeñas y de poca conectividad (Diamond & May 1976, Harris 1984, Rau & Gantz 2001), como se observa en el sector norte del PNC (Fig. 1). Por el contrario, las áreas de menor tamaño (RVS Surtubal y RVS Cacyra) mostraron cierta tendencia a ser geométricas (Cuadro 1), lo cual puede reflejar su origen privado, pues los procesos de colonización y las actividades humanas establecen linderos rectos que favorecen la aparición de formas regulares en los paisajes naturales (Rau & Gantz 2001).

El análisis del cambio de la cobertura boscosa en los tres diferentes periodos (Cuadro 2), resultó positivo, pues la cobertura forestal aumentó en 7 775,13 ha (1997 a 2000), aunque luego se redujo en 2 073,5 ha (2000-2005). Este comportamiento sugiere un aumento neto del bosque secundario en la zona de estudio, pero también indica que se ha alterado el ecosistema en algunos sitios, lo cual podría estar afectando los procesos ecológicos naturales (Bennett 1990, Guariguata & Kattán 2002). Queda pendiente el verificar *in situ* el tipo de cobertura boscosa que ha desaparecido en los últimos años, pues es de gran importancia para detener o regular las actividades de extracción maderera.

Por otro lado, las evaluaciones del hábitat en el límite oeste del PNC (carretera costanera-sur) mostraron que los cuatro puntos estudiados presentan un estado sucesional similar, lo cual puede indicar que el parche de bosque privado era parte del mismo continuo forestal del PNC, o bien que ha tenido un tiempo de regeneración natural semejante al del PNC. El hecho de que no se encontraran diferencias significativas en los valores de obstrucción horizontal (Cuadro 3) apoya ambas hipótesis; pero al comparar los valores de cobertura de dosel, se nota cierta homogeneidad dentro del PNC que no es compartida por el parche boscoso privado (Cuadro 3), lo cual sugiere que la segunda hipótesis es más apropiada.

La hipótesis de que el parche boscoso privado ha tenido un tiempo de regeneración natural semejante al del PNC también se ve reflejada en la composición arbórea del sitio (Cuadro 4), donde se observa un remanente de árboles muy grandes (DAP mayor a 40 cm) acompañados de algunos árboles menores, dando la idea de que en el pasado sufrió extracción selectiva; mientras que los sitios dentro del PNC muestran valores más homogéneos, propios de una regeneración natural que se ha mantenido por varias décadas (Vargas 1992, Vílchez *et al.* 2008).

Lo anterior es de gran importancia para la conservación de la biodiversidad, pues según observaciones de los funcionarios del PNC, el segmento de carretera entre el PNC y el parche boscoso privado es precisamente el lugar de mayor mortalidad de fauna silvestre por atropellos (A. Arce com. pers. 2011), lo cual sugiere que dicho parche boscoso le ofrece a la fauna ciertos recursos valiosos que los impulsan a desplazarse fuera del PNC con el inminente riesgo que esto conlleva (Monge-Nájera 1996, Gurrutxaga-San Vicente & Lozano-Valencia 2010).

El conservar la biodiversidad de un AP no sólo implica proteger un territorio evitando toda alteración, sino también, regular las actividades humanas aledañas para mitigar de alguna forma los efectos adversos que puedan generar hacia adentro del AP. En ese sentido, es fundamental tener presente la realidad actual del sitio y considerar que una carretera como límite del PNC produce gran afectación en la biodiversidad, no sólo por la mortalidad de la fauna silvestre, sino principalmente por impedir los desplazamientos animales (efecto de barrera) y alterar permanentemente las condiciones medioambientales a ambos lados de la carretera (efecto de borde), los cuales afectan directamente las poblaciones de vida silvestre del PNC (Arroyave *et al.* 2006).

Con miras a mantener el potencial del PNC como área de conservación de la biodiversidad, lo primero que debe recomendarse es el reducir el límite de velocidad máxima sobre la carretera costanera-sur, principalmente en la región que colinda con el PNC, pues se ha demostrado que esta simple medida reduce los atropellos considerablemente (Arroyave *et al.* 2006). Una vez reducida la mortalidad de la fauna silvestre, es importante establecer un monitoreo frecuente de la cobertura boscosa, tanto en los límites del PNC como en las AP vecinas, con el fin de establecer programas de reforestación para generar corredores biológicos funcionales que mitiguen la fragmentación, pues la zona presenta el potencial de comportarse como un continuo de bosque de considerables dimensiones, que facilitaría el establecimiento, recuperación y permanencia en el tiempo de las poblaciones de vida silvestre.

## AGRADECIMIENTOS

## LITERATURA CITADA